\documentclass[twoside]{llncs}
\usepackage{amsmath}
\usepackage{amssymb}
\usepackage{graphics}
\usepackage[dvips]{epsfig}
\usepackage{times}
\usepackage{algorithm}
\usepackage{algorithmic}
\usepackage{fancyhdr}
\usepackage{ifthen}

\usepackage[show]{ed} 

\newcommand\settitle[2][]{%
 \title{#2}
 \ifthenelse{\equal{#1}{}}%
  {\fancyhead[RO]{\nouppercase #2 \qquad \thepage}}%
  {\fancyhead[RO]{\nouppercase #1 \qquad \thepage}}%
}
\newcommand\setauthors[2]{%
 \author{#2}
  {\fancyhead[LE]{\thepage \qquad \nouppercase #1}}%
}
\fancypagestyle{plain}{%
\fancyhf{}

}

\def\keywordsname{Keywords.}
\newenvironment{keywords}{%
      \list{}{\advance\topsep by-0.50cm\relax\small
     \leftmargin=1cm
      \labelwidth=1cm
     \listparindent=1cm
     \itemindent\listparindent
      \rightmargin\leftmargin}\item[\hskip\labelsep
                                    \bfseries\keywordsname]}

   {\endlist}

   \newtheorem{Remark}{Remark}

   \def\C{{\mathbb C}}
   \def\R{{\mathbb R}}

   \def\Pf{{\it Proof.$\;\;$}}

   \def\qed{\hfill$\diamond$}
   
   \def\rk{\mbox{\rm rk~}}
   
   \def\tr{\mbox{\rm tr~}}

   \def\cV{{\mathcal V}}
   \def\cP{{\mathcal P}}

\def\C{{\mathbb C}}

\def\R{{\mathbb R}}

\def\Pf{{\it Proof.$\;$}}

\def\qed{\hspace{10cm}$\diamond$}

\def\diag{\mbox{\rm diag}}

\def\Pr{\mbox{\rm Pr}}

\def\tr{\mbox{\rm tr}}

\def\({\langle}
\def\){\rangle}
\def\mb{\boldsymbol}
\def\im{{\rm i}}

\def\cD{{\mathcal D}}
\def\cE{{\mathcal E}}

\def\cH{{\mathcal H}}

\def\cP{{\mathcal P}}

\def\cV{{\mathcal V}}
\def\cW{{\mathcal W}}

\def\1{\mb1}

\def\v0{{\bf 0}}

\def\ov{\overline}

\begin{document}

\settitle[Markov Model]{A Markovian Model for \\
Joint Observations, Bell's Inequality\\
and Hidden States}

\setauthors{U. Faigle/A. Sch\"onhuth}
           {Ulrich Faigle$^1$ and Alexander Sch\"onhuth$^2$}
\institute{Mathematisches Institut\\
           Universt\"at zu K\"oln\\
           Weyertal 80, 50931 K\"oln, Germany\\
\email{faigle@zpr.uni-koeln.de}\\[1ex]
\and
           Centrum Wiskunde \& Informatica\\
           Science Park 123\\
           1098 XG Amsterdam, The Netherlands\\
\email{A.Schoenhuth@cwi.nl}}

\date{}
\maketitle

\thispagestyle{plain}
\begin{abstract}
While the standard approach to quantum systems studies length preserving linear transformations of wave functions, the Markov picture focuses on trace preserving operators on the space of Hermitian (self-adjoint) matrices.
The Markov approach extends the standard one and provides a refined ana\-lysis of measurements and quantum Markov chains. In particular, Bell's inequality becomes structurally clear. It turns out that hidden state models are natural in the Markov context. In particular, a violation of Bell's inequality is seen to be compatible with the existence of hidden states. The Markov model moreover clarifies the role of the ''negative probabilities'' in Feynman's analysis of the EPR paradox.
\end{abstract}

\begin{keywords}
Bell's inequality, EPR paradox, hidden state, Markovian operator, negative probability, quantum Markov chain, quantum measurement
\end{keywords}
\section{Introduction \label{intro}}
Classical models of physical systems allow a clear distinction between the current state of the system and the probability to extract particular information in that state. The distinction becomes less clear in quantum models where the state of a system is given in terms of its probabilistic description {\it via} a quantum density, \emph{i.e.}, a self-adjoint matrix with trace $1$ and nonnegative eigen\-values. This situation leads to counter-intuitive difficulties that have been pointed out by Einstein, Podolsky and Rosen~\cite{Einstein}. While Feynman~\cite{Feynman} introduced the idea of ''negative probabilities'' to resolve the EPR paradox theoretically, Bell~\cite{Bell64,Bell66} offered an inequality as a test against the objection of Einstein {\it et al.}. The violation of Bell's inequality as observed experimentally by Aspect {\it et al.}~\cite{Aspect} seems to confirm the paradoxical nature of the quantum world indeed.

\medskip
We propose here a model that makes it very natural to distinguish between states and observation probabilities in quantum models. The mathematical key in our analysis is the study of general trace preserving linear operators on the space of self-adjoint matrices rather than length preserving linear operators on the space of wave functions. This approach is very much in line with the classical Markov model for evolution of states in time and thus yields a comprehensive model for quantum Markov chains as a byproduct, which includes the random walk model of Aharonov {\it et al.}~\cite{Aharonov-et-al,Kempe} as a special case, for example (Section~\ref{sec:Markov-chains}). Without going here into details, we mention that our approach also offers an alternative to Temme {\it et al.}~\cite{Temme2} for quantum Metropolis sampling.

\medskip
The elementary units in our model are Markov densities, \emph{i.e.}, self-adjoint matrices with trace $1$. Markov densities do not necessarily correspond to probabilities but are primarily mathematical descriptions of states of preparation of a system. It follows that a quantum measurement is not necessarily observable in a (macro-)statistical sense if the system is in a general Markov state. Hence it is meaningful to ask when a measurement is observable (Section~\ref{sec:Measurements}). Moreover, the Markov model yields a refined tool to analyze notions of joint observability of not just two but three or more measurements. We show in Section~\ref{sec:Bells-inequality} that the Markov perspective provides a novel look at Bell's inequality. In particular, it shows that a violation of Bell's inequality is not contradictory to but compatible with hidden state models. In fact, we argue in Section~\ref{sec:Measurement-hidden-states} that hidden state models are natural within (classical or quantum) Markov models of measurements.

\medskip
 As was pointed out previously (see \cite{deMuynck}), the EPR paradox can be reframed as resulting from EPR's stipulation that all observables associated with the preparation of a system follow a joint probability distribution which reflects a probability distribution over the inherent hidden states. Feynman suggested to resolve the problem by allowing joint probabilities to be negative, which triggers philosophical difficulties with the interpretation of such quantities. Here, we allow systems to be prepared with a clear view on their hidden states without requiring a preparation to be probabilistic. Instead, we insist that meaningful observations be statistically analyzable. So, according to the standard tenet that quantum measurements are associated with probability distributions, hidden states can perhaps not be measured statistically meaningfully. But they can very well be meaningfully encoded in terms of (Markov) densities.

\medskip
Motivated by Wigner's~\cite{Wigner} distribution, already Dirac~\cite{Dirac} and  Bartlett~\cite{Bartlett} had suggested ''negative probabilities'' to arrive at appropriate mathematical models for seemingly paradoxical quantum world situations (see also Section~\ref{sec:Feynman-EPR}) and new probabilistic models have been developed for that purpose (see, \emph{e.g.}, Khrennikov~\cite{Khrennikov}). The Markov approach clarifies the issue: Markov densities describe preparation states and not probability distributions. While those state descriptions may involve negative numbers, there are no negative probabilities. All the probability distributions that are associated with observables in the Markov model are classical (nonnegative). There is no need for new probability theories.

\section{Preliminaries}\label{sec:Preliminaries}
$\R$ denotes the scalar field of real numbers and $\C$ the field of complex numbers $z= a+\im b$ (where $a,b\in \R$ and $\im^2 =-1$). $\C^{m\times n}$ is the ($mn$)-dimensional vector space of all $(m\times n)$-matrices of the form
\begin{equation}\label{eq.matrix-form}
     C = A +\im B \quad\mbox{with $A,B\in \R^{m\times n}$.}
\end{equation}
$\ov{C} = A -\im B$ is the \emph{conjugate} of $C=A+\im B$. The transpose $C^* = \ov{C}^T$ of its conjugate $\ov{C}$ is the \emph{adjoint} of $C$. $\C^{m\times n}$ is a Hilbert space with respect to the \emph{Hermitian inner product}
$$
\(C|D\) :=  \tr(C^*D) =\sum_{i=1}^m\sum_{j=1}^n \ov{c_{ji}}d_{ij},
$$
where the $c_{ij}$ and the $d_{ij}$ denote the coefficients of $C$ and $D$. $\|C\|:= \sqrt{\(C|C\)}$ is the \emph{norm} of $C$.  $\C^n$ is short for $\C^{n\times 1}$ and can also be identified with the space of diagonal matrices in $\C^{n\times n}$:
$$
\begin{bmatrix}
v_1\\
v_2\\
\vdots \\
v_n
\end{bmatrix} \in \C^n \quad\longleftrightarrow\quad \diag(v_1,v_2,\ldots,v_n) =
\begin{bmatrix}
v_1 &0 &0 &\ldots\\
0 &v_2 &0 &\ldots\\
\vdots& &\ddots\\
0& &\ldots &v_n
\end{bmatrix}
$$

\medskip
Assuming $m=n$, a matrix $C=A+\im B$ with the property $C^*=C$ is \emph{self-adjoint} or \emph{Hermitian}, which means that $A$ is \emph{symmetric} (\emph{i.e.}, $A^T=A$) and $B$ \emph{skew-symmetric} (\emph{i.e.}, $B^T = -B$).
Let $\cH_n$ denote the collection of all Hermitian $(n\times n)$-matrices. From the general form (\ref{eq.matrix-form}) one recognizes $\cH_n$ as a real Hilbert space  of dimension $\dim_\R(\cH_n) = n^2$.

\medskip
A matrix $Q=[q_{ij}]\in \cH_n$ has real eigenvalues $\lambda_i$ and a corresponding orthonormal set $\{v_1,\ldots,v_n\}$ of eigenvectors $v_i\in \C^n$, yielding the \emph{spectral decomposition}
\begin{equation}\label{eq.Spektraldarstellung}
Q = \sum_{i=1}^n \lambda_i {v}_i v^*_i \quad\mbox{and hence trace}\quad
   \tr(Q) = \sum_{i=1}^n q_{ii} = \sum_{i=1}^n \lambda_i.
\end{equation}
$Q\in \cH_n$ is said to be \emph{nonnegative} if all eigenvalues of $Q$ are nonnegative, which is equivalent to the property
\begin{equation}\label{eq.nonnegativity}
    v^*Qv \geq 0 \quad \mbox{holds for all $v\in \C^n$.}
\end{equation}
A vector $v\in \C^n$ gives rise to a nonnegative element $vv^*\in\cH_n$ and one has
\begin{equation}\label{eq.length-trace}
  \(v|v\) = \tr(vv^*).
\end{equation}
In the case $\tr(vv^*) = 1$, the matrix $vv^*$ is thought to represent a  \emph{pure state} of an $n$-dimensional quantum system.

\medskip
A matrix $U\in \C^{n\times n\\}$ is \emph{unitary} if the identity matrix $I$ factors into $I=UU^*$, \emph{i.e.}, if the row (or column) vectors of $U$ form an orthonormal basis for $\C^n$. So also $U^*$ is unitary. For example, an orthonormal basis $\{v_1,\ldots, v_n\}$ of eigenvectors relative to $Q\in \cH_n$ gives rise to a unitary matrix $U^*$ with columns $v_i$. Where $\lambda_1,\ldots,\lambda_n$ are the corresponding eigenvalues, the linear operator $x\mapsto Qx$ on $\C^n$ is described with respect to the basis $U^*$ {\it via} the transformed matrix
\begin{equation}\label{eq.Basiswechsel}
    UQU^* = \diag(\lambda_1,\ldots,\lambda_n) \in \cH_n.
\end{equation}

\section{Measurements}\label{sec:Measurements}
Recall that a \emph{quantum measurement} with scale $\Sigma$ is a finite collection $X=\{M_a\mid a\in \Sigma\}$ of matrices $M_a\in \C^{n\times n}$ such that the self-adjoint matrices $X_a = M_aM_a^*$ yield a \emph{POVM} ($=$ positive operator valued measurement, (\emph{cf.} \cite{B-NGJ,Nielsen-Chuang})), \emph{i.e.}, the $X_a$ are non-negative and sum up to the identity:
$$
    I = \sum_{a\in \Sigma} X_a = \sum_{a\in \Sigma}M_aM_a^*.
$$

\medskip
Assume that the measurement $X$ is applied to a system $\mathfrak S$ that is in a state of preparation which is described by some $Q\in \cH_n$. If $Q$ is a quantum density, then the parameters
$$
    p_Q(a) = \(X_a| Q\) = \tr(M_aQM_a^*)
$$
are nonnegative real numbers with sum $\sum_{a\in \Sigma} p_Q(a) = \tr(Q) = 1$. In this case, $p_Q(a)$ is interpreted as the probability for the measurement $X$ to produce the value $a\in \Sigma$, which means that $X$ behaves like a random variable that can be analyzed statistically. Note, however, that $Q$ does not necessarily have to be a quantum density for $X$ be a random variable in the sense above. It suffices to have $\tr(Q) = 1$ and  $\tr(X^{(a)}Q) \geq 0$ for all $a\in \Sigma$.

\medskip
Referring to an element $Q\in \cH_n$ with $\tr(Q) = 1$ as a \emph{Markov density}, we therefore say that $X$ is \emph{(statistically) observable} relative to the Markov density $Q$ if we have
\begin{equation}\label{eq.POVM-observable}
\tr(X^{(a)}Q) \geq 0 \;\;\forall a\in \Sigma.
\end{equation}

\medskip
The notion of Markov densities $Q$ thus extends and refines the measurement model: The system $\mathfrak S$ might be in a (Markov) state of preparation that does not allow $X$ to be analyzed statistically. If $Q$ is nonnegative (and hence a quantum density), $X$ is necessarily observable in $Q$.

\subsection{Markov operators}\label{sec:Markovians}
A \emph{Markov operator} (of \emph{Markovian}) is a linear map $\mu:\cH_n\to\cH_n$ that maps Markov densities onto Markov densities. In other words: The Markovian $\mu$ is a trace preserving linear operator on $\cH_n$.
$\mu$ is said to be \emph{nonnegative} if it maps quantum densities onto quantum densities (and hence nonnegative matrices onto nonnegative matrices).

\medskip
\begin{Remark}\label{R.1} In the case of \emph{complete
  positivity} (\emph{i.e.}, $(I\otimes\mu)(A)\ge 0$ for any nonnegative
  $A\in \cW\otimes\cV$, where $\cW$ is an extra system) a Markovian operator is a \emph{quantum operation} (\emph{cf.}  \cite{Nielsen-Chuang}). The quantum operations formalism aims at modeling the dynamics of open quantum systems and quantum noise, thereby also borrowing from the interrelation between classical
  noise and classical Markov chains. Time-discrete quantum Markovian
  dynamics have also been described by trace-preserving quantum
  operations as \emph{quantum channels} (see, \emph{e.g.}, \cite{Wilde}). The more general Markovian formalism allows us to interpret quantum systems in terms of hidden states (see Section~\ref{sec:hidden-states_Bell}) and to summarize ergodic properties conveniently.
\end{Remark}

\medskip
Consider, for example, any quantum measurement $X=\{M_a\mid a\in \Sigma\}$. The associated sum operator
$$
     Q \mapsto \mu_X(Q) :=\sum_{a\in \Sigma} M_a QM_a^*
$$
is nonnegative and trace preserving and hence nonnegative Markovian. In the special case of a measurement $X_U = \{U\}$ with ''unary scale'' $\Sigma$ (\emph{i.e.}, $|\Sigma|=1$), $U$ is a unitary matrix with associated Markov operator
$$
       Q \mapsto \mu_U(Q) := UQU^*.
$$

\medskip
Any trace preserving operator $\mu:\cV\to\cV$ on some subspace $\cV\subseteq \cH_n$ extends to a trace preserving operator (and hence Markovian) $\ov{\mu}:\cH_n\to\cH_n$ in the obvious way. Consider, for example, the space $\cD$ of diagonal matrices $D\in \cH_n$. A trace preserving operator $\mu:\cD\to\cD$ is described by a
$$
    \mu(\diag(x)) = \diag(M(x)) \quad(x\in \R^n),
$$
where $M\in \R^{n\times n}$ is a \emph{Markov matrix} (\emph{i.e.}, $M$ has column sums $1$). $\mu$ is nonnegative on $\cD$ if and only if all coefficients of $M$ are nonnegative scalars, in which case $M$ is the transition matrix of a (classical) homogeneous Markov process.

\subsection{Markov measurements}\label{sec:Markov-measurements}
The quantum measurement formalism generalizes naturally. We call a family $X=\{\mu^{(a)}\mid a\in \Sigma\}$ of linear operators $\mu_a:\cH_n\to\cH_n$ a \emph{Markov measurement} with (finite) scale $\Sigma$ if
$$
   \mu_X :=\sum_{a\in \Sigma}\mu_a \quad\mbox{is a Markov operator.}
$$
$X$ is \emph{(statistically) observable} relative to the Markov density $Q$ if
$$
     \tr(\mu_a(Q)) \geq 0 \quad\mbox{holds for all $a\in \Sigma$.}
$$
Again, observability means that the measurement $X$ behaves like a random process  that produces the output $a\in \Sigma$ with probability $p_Q(a) =   \tr(\mu_a(Q))$.

\medskip
\begin{Remark}\label{R.2} When the $\mu_a$ are
  completely positive, the Markov measurement $X$ is a
  \emph{measurement model} in the sense of the quantum operations
  formalism (see \cite{Nielsen-Chuang}).
\end{Remark}

\medskip
 The application of an observable $X$ moves the system under investigation with probability $p_Q(a)$ from $Q$ into a state of preparation that is described by the Markov density $Q_a$ such that
$$
    p_Q(a) Q_a= M_a QM_a^* \quad\mbox{and hence}\quad   p_Q(a)=\tr(M_aQM_a^*).
$$
So the expected Markov density after the application of $X$ is
\begin{equation}\label{eq.Measurement-evolution}
 \sum_{a\in \Sigma} p_Q(a) Q_a  = \mu_X(Q).
\end{equation}
Multinomial expansion of the Markov measurement $X=\{\mu^{(a)}\mid a\in \Sigma\}$ yields
$$
   (\mu_X)^t = \big(\sum_{a\in \Sigma}\mu_a\big)^t = \sum_{a_1,\ldots, a_t\in \Sigma} \mu_{a_t}\mu_{a_{t-1}}\ldots \mu_{a_1} =:\sum_{w\in \Sigma^t} \mu_{w}
$$
and exhibits $X^t :=\{\mu_{w}\mid w\in \Sigma^t\}$ as a Markovian measurement with scale $\Sigma^t$ for any $t=1,2,\ldots$. We say that $X$ is \emph{$t$-observable} in $Q$ if $X^t$ is observable in $Q$. Obviously, $X$ is always $0$-observable.

\medskip
\begin{lemma}\label{l.t-observable} Assume that the Markov measurement $X$ is $t$-observable in $Q$. Then $X$ is $k$-observable in $Q$ for all $k\leq t-1$.
\end{lemma}

\Pf It suffices to consider $k=t-1$. Let $w=a_1\ldots a_{t-1}\in \Sigma^{t-1}$ be arbitrary. Since $X^t$ is observable and $\mu_X$ Markovian, one has
$$
\tr(\mu_w(Q)) = \tr(\mu_X(\mu_w(Q))) =\sum_{a\in \Sigma}\tr(\mu_{wa}(Q)) \geq 0.
$$
\qed

\medskip
If $X$ is $k$-observable in $Q$, the application of $X^k$ yields the expected Markov density $\mu^k_X(Q)$. The following is straightforward.

\medskip
\begin{lemma}\label{l.t-evolution} Assume that the Markov measurement $X$ is $t$-observable in $Q$. Then $X$ is observable in $\mu^k(Q)$ for all $k\leq t-1$.

\qed
\end{lemma}

\medskip
We say that $X$ is \emph{completely observable} in $Q$ if $X$ is $t$-observable in $Q$ for all $t\geq 1$.

\medskip
\begin{Remark}\label{R.3} With the methods of \cite{FS} one can show that completely observable Markov measurements $X$ yield asymptotically mean stationary processes.  Such processes are known to have good ergodic properties (see \cite{Gray,Gray1}). For example, the conditional entropies converge to a limit
$$
H_\infty(X)= \lim_{t\to \infty}\frac1t H(X^t) = \lim_{t\to\infty} H(X^t|X^{t-1}).
$$
In the case of diagonal matrices, complete observability is equivalent to the much studied concept of \emph{finite-dimensional} (classical) stochastic processes or \emph{observable operator models} that provide a proper generalization of classical (possibly hidden) discrete Markov chains with finite alphabets  (see \cite{Gilbert,Ito,Jaeger}). We are lead to the latter model in more detail in Section~\ref{sec:Measurement-hidden-states}.
\end{Remark}

\section{Markov chains}\label{sec:Markov-chains}
Let $\mu$ be a Markov operator on $\cH_n$. Then any Markov density $P$ gives rise to a chain of Markov densities $\mu^t(P)$,
$$
(\mu,P) := \{\mu^t(P)\mid t=0,1,\ldots \},
$$
where $\mu^0$ is the identity operator on $\cH_n$. We refer to $\mu$ as the \emph{evolution operator} of the \emph{Markov chain} $(\mu,P)$. In the case $\mu(P) = P$, the chain $(\mu,P)$ is \emph{stationary}. Note that classical Markov chains correspond to Markov chains of nonnegative diagonal matrices in the general model.

\medskip
\begin{Remark}\label{R.4}
The Markovian evolution operator $\mu$ does not have any density or probability interpretation in its own right but simply describes how the densities in a chain $(\mu,P)$ evolve over time.
\end{Remark}

\medskip
The Markov chain $(\mu,P)$ is \emph{bounded} if there exists some $c\in \R$ such that
$$
    \(\mu^t(P)|\mu^t(P)\) = \tr(\mu^t(P)^2) \leq c \quad\mbox{holds for all $t$.}
$$
Assume, for example, that $\mu$ in nonnegative and $P$ a quantum density. Then the Markov chain $(\mu,P)$ is bounded because each $\mu^t(P)$ is a quantum density and therefore satisfies $\tr(\mu^t(P)^2)\leq 1$.

\medskip
{\bf Evolution of wave functions.} Any unitary matrix $U\in \C^{n\times n}$ induces a nonnegative Markovian $\mu_U(C) = UCU^*$. On the other hand, a wave function $v\in \C^n$ of norm $\|v\|=1$ is transformed under $U$ into the wave function $Uv$ of norm $\|Uv\|=1$, which corresponds to the pure quantum state
$$
 (Uv)(Uv)^* =  Uvv^*U^* =  \mu_U(vv^*).
$$
So $(\mu_U, vv^*)$ is the Markov chain of the pure quantum states that correspond to the wave function evolution $v\mapsto U^tv$ for $t=0,1,\ldots$.

\medskip
{\bf Quantum random walks on graphs.} A model for a quantum random walk on a directed regular graph $G$ has been proposed by Aharonov
\emph{et al.}~\cite{Aharonov-et-al} (see also Kempe~\cite{Kempe}). A natural generalization of that model to arbitrary directed graphs on $V$ is the following.

\medskip
Given the graph $G=(V,\cE)$ with node set $V$ and edge set $\cE\subseteq V\times V$, we identify the elements  $e\in \cE$ with the elements in the standard basis of unit vectors in $\C^n$. A quantum walk on $G$ is now described in terms of a Markovian operator $\mu:\cH_n\to\cH_n$. The walk on $G$ starts at time $t=0$ in a quantum density state $P^{(0)}$ and consists in subsequent density changes according to
$$
        P^{(t)} := \mu(P^{(t-1)}) \quad\mbox{for all \;$t=1,2,\ldots$}.
$$
  The quantum walk corresponds to the Markov chain $(\mu,P^{(0)})$ in our terminology. Let us define for every node $v\in V$ the diagonal matrix $X_v = \diag(x^{(v)}) \in \cH_n$, where the vector $x^{(v)}\in\C^n$ has the components
$$
     x^{(v)}_e := \left\{\begin{array}{cl} 1 &\mbox{if $e\in \cE$ is of the form $e=(v,w)$}\\
     0 &\mbox{otherwise.}\end{array}\right.
$$
Note that the family $X=\{X_v\mid v\in V\}$ constitutes a POVM. If all $P^{(t)}$ are nonnegative,  the application of the matrices $X^{(v)}$ yields the nonnegative parameters
$$
p_t(v) = \(X_v|P^{(t)}\) = \tr(X_v P^{(t)})\geq 0,
$$
which means that $X$ is observable. $p_t(v)$ is then interpreted as the probability for the random walk to be in node $v$ at time $t$.

\medskip
Aharonov \emph{et al.}~\cite{Aharonov-et-al} assume a unitary wave function evolution and hence a non\-negative Markovian operator of the type $\mu_U(P) = UPU^*$. Using the assumed regularity of the underlying graph $G$, they derive an ergodic property of their quantum walk and show that the following limit of averages exists in their model:
$$
     \ov{p}(v) =  \lim_{t\to \infty}
     \frac1t\sum_{k=0}^{t-1} p_k(v) \;.
$$
We will exhibit a general convergence property for bounded Markov chains below.

\medskip
\begin{Remark}\label{R.2} Our general quantum random walks need not respect the structure of the underlying graph. This is purely for formal convenience: Our model incorporates the most general graph-respecting concepts of quantum walks suggested in \cite{Aharonov-et-al}.
\end{Remark}

\medskip
It is of interest to know whether a Markov chain converges towards a stationary behavior. We consider a fixed Markovian $\mu:\cH_n\to\cH_n$.

\medskip
\begin{theorem}\label{t.Markov-limit} Let $(\mu,P)$  be a bounded Markov chain. Then the limit of averages
\begin{equation}\label{eq.Cesaro-limit}
   \tilde{P} = \lim_{t\to\infty}\frac1t\sum_{k=1}^t \mu^t(P)
\end{equation}
exists and $(\mu,\tilde{P})$ is a stationary Markov chain. Moreover, if $\mu$ is nonnegative and $P$ a quantum density, then also $\tilde{P}$ is a quantum density. \end{theorem}

\medskip
If the limit (\ref{eq.Cesaro-limit}) exists, it is clear that $\tr(\tilde{P})=\tr(P)=1$ holds and $(\mu,\tilde{P})$ is stationary. Moreover, if $\mu$ preserves quantum densities, then each $\mu^t(P)$, and therefore each average, is a quantum density. So it remains to prove the existence of $\tilde{P}$. It is convenient to base the proof on the following lemma.

\medskip
\begin{lemma}[\cite{FS}]\label{l.FS} Let $V$
be a finite-dimensional normed vector space over $\C$ and consider
the linear operator $F:V\to V$. The following statements are
equivalent:
\begin{itemize}
\item[(a)] $\ov{v} = \lim_{t\to\infty}\frac1t\sum_{k=1}^{t-1}F^k(v)$
exists for all $v\in V$.
\item[(b)] For every $v\in V$, there exists some finite bound $c\in \R$ such that
$\|F^t(v)\|\leq c$ holds for all $t\geq 0$.
\end{itemize}
\end{lemma}

\medskip
We want to apply Lemma~\ref{l.FS} to the (complex) subspace $V\subseteq \C^{n\times n}$ that is generated by $(\mu,P)$ with the norm $\|C\| =\sqrt{\tr(C^*C)}$. To this end, we augment $P$ to a basis $\{Q_0,Q_1,\ldots,Q_m\}\subseteq (\mu,P)$ for $V$ with $Q_0=P$ and define the linear operator $F:V\to V$ {\it via}
$$
     F\big(\sum_{j=0}^m r_j Q_j\big) := \sum_{j=0}^m r_j\mu(Q_j).
$$
Then $F$ agrees with $\mu$ on the Markov chain $(\mu,P)$. Let $c_1$ be a bound on $(\mu,P)$ and observe from the triangle inequality:
$$
\|F^t\big(\sum_{i=1}^m r_i Q_i\big)\| \;\leq\; \sum_{i=1}^m
|r_i|\cdot\|\mu^t(Q_i)\| \;\leq\; \sum_{i=1}^m |r_i|c_1 =:c\;.
$$
So $F$ satisfies condition (b) and
hence also (a) of Lemma~\ref{l.FS}, which establishes the convergence of the averages in Theorem~\ref{t.Markov-limit} with the choice $v=P$.

\medskip
\begin{corollary}\label{c.convergence} Let $(\mu,P)$ be a bounded Markov chain and $X:\cH_n\to \R$ any linear functional. Then
$$
\lim_{t\to\infty}\frac1t\sum_{k=1}^t X(\mu^t(P))  = X(\tilde{P}).
$$
\qed
\end{corollary}

\medskip
\begin{Remark}\label{R.6} In contrast to density averages, wave function averages in a wave function evolution will generally \emph{not} converge to a wave function. If $U$ is a unitary matrix that does not have $1$ as an eigenvalue, one has
$$
\lim_{t\to\infty} \frac1t\sum_{k=1}^t U^tv  = 0 \quad\mbox{for all $v\in \C^n$.}
$$
\end{Remark}

\section{Hidden states and Bell's inequality}\label{sec:hidden-states_Bell}
Let $\mathfrak{S}$ be a physical system with a finite set $\Omega=\{\omega_1,\ldots,\omega_N\}$ of \emph{hidden states} and assume that $\mathfrak S$ is (definitely) in one of the $N$ possible hidden states $\omega\in \Omega$ at any discrete time $t=0,1,\ldots$. It may not be possible to observe the true hidden state $\omega=\omega^{(t)}$ at time $t$ directly. So we want to collect statistical information on the true state.

\subsection{Information functions}\label{sec:Information-functions} An \emph{information function} with scale $\Sigma$ is a function $X:\Omega\to \Sigma$. Since $\Omega$ is finite, we may assume $\Sigma$ to be  finite as well. Equivalently, $X$ can be viewed as a POVM $\{X_a\mid a\in \Sigma\}$ of diagonal matrices $X_a=\diag(x^{(a)})$, with $x^{(a)}_{\omega}= 1$ if $X(\omega) = a$.  
Similar to our terminology in Section~\ref{sec:Measurements}, we call a vector $q\in \R^\Omega$ with components $q_\omega$ a \emph{Markov state} of $\mathfrak S$ if
$$
      \sum_{\omega\in \Omega} q_\omega = 1.
$$
For any $a\in \Sigma$, we set $$
    p_q(a) := \sum_{\omega:X(\omega)=a} q_\omega
$$
and say that the information function $X$ is \emph{(statistically) observable} in $q$ if all the $p_q(a)$ are nonnegative numbers. In the case of observability and a real-valued information function $X$, we obtain the well-defined expectation
$$
E_q(X) := \sum_{x\in \Sigma} x\cdot p_q(x) = \sum_{\omega\in \Omega} X(\omega)q_\omega.
$$

\subsection{Joint observations}\label{sec:Joint-observations}
 We say that the $k$ information functions  $X_1:\Omega\to \Sigma_1, \ldots, X_k:\Omega \to \Sigma_k$ on the system $\mathfrak S$ are  \emph{jointly observable} in the Markov state $q$ if the composite information function $X:\Omega\to\Sigma$ with
$$
   X(\omega) := (X_1(\omega),\ldots, X_k(\omega))\quad\mbox{and}\quad \Sigma :=\Sigma_1\times \ldots \times \Sigma_k
$$
is observable in $q$. The following statement is easy to verify.

\medskip
\begin{lemma}\label{l.joint-observations} Assume that the collection of $k$ information functions $X_1,\ldots, X_k$ is jointly observable in the Markov state $q$, then every subcollection $X_{i_1},\ldots, X_{i_m}$ is jointly observable in $q$. In particular, every individual information function $X_i$ is observable.
Moreover, if the $X_i$ are real-valued, also every product $X_iX_j$ is observable in $q$.

\qed
\end{lemma}

\medskip
Hence, if two information functions $X$ and $Y$ on the system $\mathfrak S$ are real-valued and jointly observable, their product $XY$ is statistically observable and has a well-defined expectation $E(XY)$.

\medskip
Clearly, any collection of information functions is jointly observable in any non\-negative Markov  state $q$, \emph{i.e.}, classical probability distribution $q$ on $\Omega$.

\subsection{Feynman's approach to the EPR paradox}\label{sec:Feynman-EPR} Feynman~\cite{Feynman} has given a mathematical model to explain the Einstein, Rosen and Podolsky (EPR) paradox (see also Scully {\it et al.}~\cite{Scully-Walther-Schleich}). In essence, Feynman provides the example of two information functions $X,Z$ for the spin along the $+x$ and $+z$ axis of a spin $1/2$ system $\mathfrak S$ with $\Omega=\{(++),(+-),(-+),(--)\}$ as in the following table:
$$
\begin{array}{c|cccccc}
 &(++)&(+-)&(-+)&(--) \\  \hline
X &+&+&-&-\\
Z &+&-&+&-
\end{array}
$$
and the ''joint probability distribution'':
\begin{eqnarray*}
P(++) &=& [1+\(\hat{\sigma}_z\) + \(\hat{\sigma}_x\)+\(\hat{\sigma}_y\)]/4\\
P(+-) &=& [1+\(\hat{\sigma}_z\) - \(\hat{\sigma}_x\)-\(\hat{\sigma}_y\)]/4\\
P(-+) &=& [1+\(\hat{\sigma}_z\) + \(\hat{\sigma}_x\)-\(\hat{\sigma}_y\)]/4\\
P(--) &=& [1-\(\hat{\sigma}_z\) - \(\hat{\sigma}_x\)-\(\hat{\sigma}_y\)]/4,
\end{eqnarray*}
where $\(\hat{\sigma}_x\),\(\hat{\sigma}_y\),\(\hat{\sigma}_z\)$ are the Pauli spin operators. Noticing that those parameters $P(xz)$  might take on negative values, Feynman interprets them as (possibly) \emph{negative} probabilities and argues that probability theory should be expanded in this direction.

\medskip
In our terminology, the $P(xz)$ yield a blueprint for attainable Markov states (depending on the actual values of the spin operators), where $X$ and $Z$ are not always guaranteed to be jointly statistically observable. For example, the situation $$
\(\hat{\sigma}_x\)=\(\hat{\sigma}_y\)=\(\hat{\sigma}_z\)= 1/2,
$$
would result in the Markov state $q=(5/8, 1/8, 3/8, -1/8)$, where $X$ and $Z$ are individually but not jointly observable.

\subsection{Bell's inequality}\label{sec:Bells-inequality}
The well-known inequality of Bell~\cite{Bell64,Bell66} takes the form (\ref{eq.Bell}) in our context as a statement on the expectations of products of pairs of information functions.

\medskip
\begin{lemma}[Bell's inequality]\label{l.Bell} Let $X,Y,Z:\Omega\to\{-1,+1\}$ be
arbitrary information functions on the system $\mathfrak S$. If $X,Y$ and $Z$ are jointly observable in the Markov state $q$, then the following inequality holds:
\begin{equation}\label{eq.Bell}
|E_q(XY) -E_q(YZ)| \;\leq\; 1 - E_q (XZ) \;.
\end{equation}
\end{lemma}

{\small\medskip \Pf Any choice of $x,y,z\in \{-1,+1\}$ satisfies the
inequality $|xy -yz| \;\leq\; 1-xz$. Because of the joint observability
assumption, all the observation probabilities
$$
p_q(x,y,z)= \Pr\{X=x,Y=y,Z=z\}
$$
are nonnegative real numbers that sum up to $1$. So we conclude
\begin{eqnarray*}
|E_q(XY) -E_q(YZ)| &=&\big|\sum_{x,y,z}
(xy-yz)p_q(x,y,z)\big|
\;\leq\; \sum_{x,y,z}|xy-yz| p_q(x,y,z)\\
&\leq& \sum_{x,y,z}(1-xz) p_q(x,y,z)\;=\; 1 -E_q(XZ)\;.
\end{eqnarray*}

\qed}

\medskip
Bell's inequality may be violated by information functions that are pairwise but not jointly observable. Consider a system $\mathfrak S$ with a set $\Omega=\{\omega_1,\omega_2,\omega_3,\omega_4,\omega_5\}$ of five hidden states, for example, and three information functions $X,Y,Z:\Omega \to \{-1,+1\}$ as in the following table:
$$
\begin{array}{c|cccccc}
 &\omega_1&\omega_2&\omega_3&\omega_4&\omega_5 \\  \hline
X &-1&+1&-1&-1&-1 \\
Y &+1&+1&-1&+1&-1 \\
Z  &+1&+1&+1&-1&-1
\end{array}
$$
One can check that $X,Y,Z$ are pairwise observable in the Markov state
$$
      q=(-1/3,1/3,1/3,1/3,1/3)
$$
and yield the product expectations
$$
   E_q(XY) = +1,\; E_q(YZ) = -1/3,\; E_q(XZ)= +1\;,
$$
which violate Bell's inequality (\ref{eq.Bell}).

\medskip
\begin{Remark}\label{R.7} Experimental results seem to indicate that quantum systems may violate Bell's inequality (see, \emph{e.g.}, Aspect \emph{et al.}~\cite{Aspect}). This is sometimes interpreted as showing that quantum mechanics does not admit a theory with hidden variables. The Markovian picture
 makes it clear that a violation of Bell's inequality only shows that the system is studied in terms of measurements that are perhaps pairwise but not jointly observable. The existence of definite but hidden states is not excluded.
In fact, an experimentally observed violation of Bell's inequality suggests that one should not place \emph{a priori} nonnegativity restrictions on concepts of states into which a system can be prepared.
\end{Remark}

\section{Measurement processes and hidden states}\label{sec:Measurement-hidden-states}
Let us now argue that statistical models for the analysis of measurements involve Markov states with possibly negative components quite naturally.

\medskip
 Suppose that some measurement $X$ with finite scale $\Sigma$ is made on some (not further specified) system $\mathfrak S$ at discrete times $t=1,2,\ldots$. Let $X_t$ be the result observed at time $t$ and assume that the results of the measurements are jointly observable in the sense that the $X_t$ are random variables with joint probabilities
 $$
   p(a_1\ldots a_t) = \Pr\{X_1=a_1,\ldots, X_t=a_t\} \quad\mbox{for all $a_1,\ldots,a_t\in \Sigma$.}
$$
For any $w=a_1\ldots a_t$ and $v= a_{t+1}\ldots a_{t+s}$, the conditional probability
$$
    p(v|w) = \Pr\{X_{t+1} = a_{t+1},\ldots, X_{t+s} = a_{t+s}| X_1=a_1,\ldots,X_t=a_t\}
$$
is the probability for correctly predicting the sequence $v$ to occur in the next $s$ observations when $w$ is obtained up to time $t$. (For formal reasons, it is convenient to assume that the \emph{empty symbol} $\Box$ is issued at time $t=0$ with probability $p(\Box) = 1$.) We collect all the prediction probabilities into the (infinite) \emph{prediction matrix}
$$
     \cP = [p(v|w)].
$$
$\cP$ is a Hankel matrix. The rank $\rk(\cP)$ is also known as the \emph{dimension} of $X$. For example, classical (possibly hidden) Markov chains with finite alphabet $\Sigma$ can be shown to be finite-dimensional (\cite{Gilbert,Ito,Jaeger}). Moreover, finite-dimensional processes are  asymptotically mean stationary (see \cite{FS}).

\medskip
In the case $m=\rk(\cP)<\infty$ it is not difficult to see (\emph{cf.} \cite{Jaeger}) that $(X_t)$ admits an \emph{observable operator model} in the following sense:
\begin{itemize}
\item[$\bullet$] There is a family $X=\{M_a|a\in \Sigma\}$ of real ($m\times m$)-matrices $M_a$ and a\\ Markov state  $\pi\in \R^m$ such that for all $a_1,\ldots, a_t\in \Sigma$
\begin{itemize}
\item[(1)] $p(a_1\ldots a_t)=\tr(\diag(M_{a_t}M_{a_{t-1}}\ldots M_{a_1}\pi))$;
\item[(2)] $M=\sum_{a\in \Sigma} M_a$ is a Markov matrix.
\end{itemize}
\end{itemize}

In our terminology, this means that $X$ represents a Markov measurement with respect to diagonal matrices that is completely observable relative to the Markov density $\Pi=\diag(\pi)$.

\medskip
\begin{Remark}\label{R.8}
Finite-dimensional measurement processes that do \emph{not} admit an observable operator model with only nonnegative Markov states are known to exist (\emph{cf.} \cite{Jaeger}).
\end{Remark}

\medskip
Let $(X_t)$ be a measurement process on a system $\mathfrak S$ with finite alphabet $\Sigma$ and finite dimension $m$ and the associated observable operator model $(\{M_{a}|a\in \Sigma\}, \pi)$. Consider
$$
   \Omega = \{(a,j)| a\in \Sigma, j=1,\ldots,m\}
$$
as the set of $N =|\Sigma|\cdot m$ hidden states of $\mathfrak S$ with the information function
$$
      X:\Omega \to \Sigma \quad\mbox{such that}\quad X(a,j)= a.
$$
With each $M_{a}=[m^{(a)}_{ij}]\in \R^{m\times m}$, we associate the
$(N\times N)$-matrix
$$
\ov{M}_{a} = [m_{(b,i),(b,j)}]\quad\mbox{with}\quad
m_{(b,i),(c,j)} = \begin{cases} m^{(a)}_{ij} &\text{if $b=a$}\\
   0 &\text{if $b\neq a$.}\end{cases}
$$
Let furthermore $\ov{\pi}$ be the $N$-dimensional vector with coordinates
$$
  \ov{\pi}(a,j) = \pi(j)/|\Sigma|.
$$
By construction, we have now for all $w=a_1\ldots a_t$,
$$
    p(w) = \tr(\diag(\ov{M}_{w}\ov{\pi}))  \;.
$$

The process $(X_t)$ can thus be interpreted as emanating from
$\mathfrak S$  {\it via} the information function $X$. The system is prepared to be in the Markov state $\ov{\pi}$ at time $t=0$. At time $t\geq 1$, the system is in a hidden state $\omega\in \Omega$ yielding the information $X(\omega) = a$ with probability
$$
 \Pr\{X_t=a\} =  \sum_{\omega: X(\omega) = a} \ov{\pi}^{(t)}(\omega)\quad
 \mbox{(where $\ov{\pi}^{(t)}= \ov{M}^t\ov{\pi}$).}
$$

\bibliographystyle{splncs}

\end{document}